\documentclass[proceedings]{stacs}
\stacsheading{2009}{147--158}{Freiburg}
\firstpageno{147}

\usepackage{amsmath}
\usepackage{amssymb}

\theoremstyle{plain}\newtheorem{question}{Question}

\newcommand{\cs}{2^\omega}
\newcommand{\fs}{2^{<\omega}}
\newcommand{\N}{\mathbb{N}}

\newcommand{\kub}{\mathfrak{K}}
\newcommand{\U}{\mathbb{U}}
\newcommand{\uh}{\upharpoonright}

\begin{document}

\title[Kolmogorov complexity and Solovay functions]{Kolmogorov complexity and Solovay functions}

\author[lab1]{L. Bienvenu}{Laurent Bienvenu}
\author[lab1]{R. Downey}{Rod Downey}
\address[lab1]{School of Mathematics, Statistics and Computer Science
  \newline Victoria University
  \newline P.O. Box 600
  \newline Wellington, New Zealand}  
\email{{laurent.bienvenu,rod.downey}@mcs.vuw.ac.nz} 




\thanks{The authors are supported by a grant from the Marsden fund of New Zealand} 



\begin{abstract}
  \noindent Solovay~\cite{Solovay1975} proved that there exists a computable upper bound~$f$ of the prefix-free Kolmogorov complexity function~$K$ such that $f(x)=K(x)$ for infinitely many~$x$. In this paper, we consider the class of computable functions~$f$ such that $K(x) \leq f(x)+O(1)$ for all~$x$ and $f(x) \leq K(x)+O(1)$ for infinitely many~$x$, which we call Solovay functions. We show that Solovay functions present interesting connections with randomness notions such as Martin-L\"of randomness and K-triviality.    
 \end{abstract}

\maketitle

\section{Introduction}\label{sec:intro}

The plain and prefix-free Kolmogorov complexities (which we denote respectively by~$C$ and~$K$) are both non-computable functions, but they do admit computable \emph{upper bounds}. How good can these upper bounds be? That is, how close to~$C$ (resp.\ $K$) can a computable upper bound of~$C$ (resp.\ $K$) be? It can be easily proven that no  computable upper bound of $C$ can be close to $C$ on all values, i.e.\ given any computable upper bound~$f$ of~$C$, the ratio $f(x)/C(x)$ is not bounded. To see this, we use a variation of Berry's paradox: take a computable upper bound~$f$ of $C$, and define, for all~$n \in \N$, $x_n$ to be the smallest string $x$ such that $f(x) \geq n$. Since~$f$ is computable, $x_n$ can be computed from~$n$, hence $C(x_n) \leq \log(n)+O(1)$. Thus, $f(x_n)/C(x_n) \geq n/(\log(n)+O(1))$ which proves the result. The exact same argument shows that no computable upper bound of~$K$ approximates~$K$ well on all values.\\

Therefore, one may ask the natural question: are there computable upper bounds for $C$ or $K$ that are good approximations on infinitely many values? The answer is trivially yes for~$C$. Indeed, for most strings~$x$, we have $C(x)=|x|+O(1)$ (see for example Downey and Hirschfeldt~\cite{DowneyHirschfeldtTA}), hence for some constant~$c$, the function~$f$ defined by $f(x)=|x|+c$ is a computable upper bound of $C$, and $f(x)=C(x)+O(1)$ for infinitely many strings~$x$. The case of~$K$ is less clear: indeed, the maximal prefix-free complexity of a string $x$ of length~$n$ (attained by most strings of that length) is $n+K(n)+O(1)$. And giving a good upper bound of this last expression already necessitates a good upper bound on~$K$! Solovay~\cite{Solovay1975} nonetheless managed to construct a computable upper bound~$f$ of~$K$ such that $f(x)=K(x)$ for infinitely many~$x$. In this paper, we consider the class of computable functions~$f$ such that $K(n) \leq f(n)+O(1)$ for all~$n$ and $f(n) \leq K(n)+O(1)$ for infinitely many~$n$, which we call Solovay functions.\\

Our first main result (Theorem~\ref{thm:mlr-sum}) is that Solovay functions have a very simple characterization: they correspond to the computable functions $f$ such that $\sum_x 2^{-f(x)}$ is finite and is a Martin-L\"of random real. 

Then, we discuss the role of Solovay functions in the characterization of randomness notions. In particular, we show (Theorem~\ref{thm:miller-yu-solf}) that Solovay functions are particularly relevant to the Miller-Yu characterization of Martin-L\"of random sequences via the plain Kolmogorov complexity of the initial segments. We prove along the way a theorem of independent interest (Theorem~\ref{thm:no-gap}) showing that the Levin-Schnorr characterization of Martin-L\"of randomness by prefix-free Kolmogorov complexity is very sharp, and derive several interesting consequences of this result.

Finally, we study two triviality notions that relate to computable upper bounds of prefix-free Kolmogorov complexity and Solovay functions. In the spirit of the Miller-Yu theorem, we obtain (Theorem~\ref{thm:k-triv-kub}) a characterization of K-triviality via computable upper bounds of~$K$.\\

We assume that the reader is familiar with the field of algorithmic randomness. If not, one can consult Downey and Hirschfeldt~\cite{DowneyHirschfeldtTA} or Nies~\cite{Nies2009}. We denote by $\fs$ and $\cs$ the set of binary sequences (or ``strings'') and binary infinite sequences respectively. For a binary sequence $x$ (finite or infinite), we denote by $x(i)$ the $(i+1)$-th bit of~$x$, and by $x \uh i$ the string made of the first~$i$ bits of~$x$ (that is, $x \uh i=x(0)x(1)\ldots x(i-1)$). The length of a string $x$ is denoted by $|x|$. Throughout this paper, we identify $\fs$ with $\N$, via the usual length-lexicographic bijection: $0=\epsilon$ ($\epsilon$~being the empty string), $1=0$, $2=1$, $3=00$, $4=01$, $5=10$\ldots. We also identify any element $r \in [0,1]$ to an element $\alpha \in \cs$ such that $r=\sum_n  \alpha(n)2^{-n+1}$. If $r$ is not dyadic 
then $\alpha$ is unique; if $r$ is dyadic, there are two possible choices for $\alpha \in \cs$ and which one we choose does not matter in this paper. We say that a real number is left-c.e.\ if it is the limit of a  computable nondecreasing sequence of rational numbers. Given a nondrecreasing unbounded function $f:\N \rightarrow \N$, we denote by $f^{-1}$ the function defined by $f^{-1}(k)=\min\{n \in \N \mid f(n) \geq k\}$.  As we stated earlier, we denote by $C(x)$ and $K(x)$ the plain Kolmogorov complexity and prefix-free Kolmogorov of a string~$x$. Since~$C$ and~$K$ are enumerable from above (i.e.\ their upper graph is a c.e.\ set), for a fixed enumeration, let $C_s(x)$ and $K_s(x)$ be the value of $C(x)$ and $K(x)$ at the $s$-th stage of the enumeration. In particular, this means that the function $(x,s) \mapsto K_s(x)$ is computable and, for any fixed~$x$, $s \mapsto K_s(x)$ is nonincreasing and converges to~$K(x)$ (and the same is true for~$C$).

\section{Computable upper bounds of Kolmogorov complexity}

The class of computable upper bounds of Kolmogorov complexity has been studied in Bienvenu and Merkle~\cite{BienvenuM2007b} (in the setting of ``decidable machines''), where they were used to give  characterizations of a wide variety of randomness notions of randomness, such as Martin-L\"of randomness, Schnorr randomness, Kurtz randomness or computable dimension. Here, we are interested in the class of Solovay functions, which is a subclass of computable upper bounds of~$K$ (here and from now on, we use a slight abuse of terminology, calling ``upper bound'' of $K$ a function~$f$ such that $K \leq f+O(1)$). 

Let us first mention that computable upper bounds of $K$ admit a very simple characterization.

\begin{lem}\label{lem:kub-sum}
Let~$f:\N \rightarrow \N$ be a computable function. The following are equivalent:\\ 
(i) $K \leq f+O(1)$\\
(ii) The sum $\sum_{n \in \N} 2^{-f(n)}$ is finite.   
\end{lem}

\begin{proof}
$(i) \Rightarrow (ii)$ is trivial as $\sum_n 2^{-K(n)} \leq 1$. For $(ii) \Rightarrow (i)$, let $c$ be such that $\sum_n 2^{-f(n)} \leq 2^c$. Using the Kraft-Chaitin theorem, we effectively construct a prefix-free c.e.\ set of strings $\{x_n \mid n \in \N\}$ with $|x_n|=f(n)+c$ for all~$n$. Then, we define a (computable) function~$F$ by $F(x_n)=n$ for all~$n$. Since $F$ has prefix-free domain and is computable, we have $K(n) \leq |x_n|+O(1)$ for all~$n$, hence $K(n) \leq f(n)+c+O(1)$.  
\end{proof}

\begin{defi}
We denote by $\kub$ the class of computable functions such that {${\sum_n 2^{-f(n)} < +\infty}$} (or equivalently, the computable functions~$f$ such that $K \leq f+O(1)$).\\
We call \emph{Solovay function} any function~$f \in \kub$ such that $\liminf_{n \rightarrow +\infty}\, f(n)-K(n) < +\infty$ (or equivalently, any function $f \in \kub$ such that for some~$c$, $f(n) \leq K(n)+c$ for infinitely many~$n$). 
\end{defi}

\begin{thm}[Solovay~\cite{Solovay1975}]\label{thm:existence-solf}
Solovay functions exist.
\end{thm}

\begin{proof}
Let us start by an observation. Given $x \in \fs$, and some~$p$ such that $\U(p)=x$, if we call~$t$ the computation time of $\U(p)$, we have $$K\left( \langle x,p,t \rangle \right) \leq |p|+O(1)$$
(where $\langle .,.,. \rangle$ is a computable bijection from $\fs \times \fs \times \fs$ to $\fs$). Indeed, given~$p$ only, one can easily compute $x$ and $t$. Suppose now that $p$ is a \emph{shortest} $\U$-program for~$x$  i.e.\ $\U(p)=x$ and $K(x)=|p|$. We then have:
$$
|p| = K(x) \leq K\left( \langle x,p,t \rangle \right) \leq |p|+O(1)
$$
Thus, let~$f$ be the function defined by:
$$
f(\langle x,p,t \rangle)=\left\{ \begin{array}{l} |p|~\text{if $\U(p)$ outputs~$x$ in exactly~$t$ steps of computation} \\ +\infty~\text{otherwise} \end{array}\right.
$$
(here we use the value $+\infty$ for convenience, but any coarse upper bound of $K(\langle x,p,t \rangle)$, like $2|x|+2|p|+2\log t$, would do). 
By the above discussion, we have $K \leq f+O(1)$ and $f(\langle x,p,t \rangle) \leq K(\langle x,p,t \rangle)+O(1)$ for all triples $(x,p,t)$ such that $p$ is a shortest $\U$-program for~$x$ and $\U(p)$ outputs~$x$ in exactly~$t$ steps of computation. Thus, $f$ is as desired. \end{proof}

\begin{remark}
In fact, what Solovay actually proved is that there exists a computable function~$f$ such that $K \leq f$ and $K(n)=f(n)$ for infinitely many~$n$. This can be easily deduced from Theorem~\ref{thm:existence-solf}. Indeed, given a computable function $f$ such that $K \leq f+O(1)$ and $c=\liminf_{n \rightarrow +\infty}\, f(n)-K(n) < +\infty$, 
the (computable) function $f'=f-c$ is such that $f'(n)=K(n)$ for infinitely many~$n$, and $f'(n) \geq K(n)$ for almost all~$n$. Hence, up to modifying only finitely many values of $f'$, we may assume that $f'(n) \geq K(n)$ for all~$n$.  
\end{remark}

It turns out that, among the computable functions~$f$ such that the sum $\sum_n 2^{-f(n)}$ is finite, the Solovay functions are precisely those for which this sum is not only finite but also a Martin-L\"of random real.

\begin{thm}\label{thm:mlr-sum}
Let $f$ be a computable function. The following are equivalent:\\ 
(i) $f$ is a Solovay function.\\
(ii) The sum $\sum_n 2^{-f(n)}$ is finite and is a Martin-L\"of random real.
\end{thm}

\begin{proof}
$(i) \Rightarrow (ii)$. If $f$ is a Solovay function, we already know by definition that $\alpha=\sum_n 2^{-f(n)}$ is finite. Let us now prove that $\alpha$ is a Martin-L\"of random real. Suppose it is not. Then for arbitrarily large~$c$ there exists~$k$ such that $K(\alpha \uh k) \leq k-c$ (this because of the Levin-Schnorr theorem, see next section). Given $\alpha \uh k$, one can effectively find some~$s$ such that
$$
\sum_{n >s} 2^{-f(n)} \leq 2^{-k}
$$
Thus, by a standard Kraft-Chaitin argument, one has $K(n| \alpha \uh k) \leq f(n)-k+O(1)$ for all $n>s$. Thus, for all $n>s$:
$$
K(n) \leq f(n)+K(\alpha \uh k)-k+O(1) \leq f(n)+(k-c)-k-O(1) \leq f(n)-c-O(1)
$$
And since~$c$ can be taken arbitrarily large, this shows that $\lim_{n \rightarrow +\infty} f(n)-K(n) = +\infty$ i.e.\ $f$ is not a Solovay function.\\

\noindent $(ii) \Rightarrow (i)$. Suppose now for the sake of contradiction that~$f$ is not  a Solovay function and that $\alpha$ is Martin-L\"of random. We will prove that under these assumptions, the number $\Omega=\sum_n 2^{-K(n)}$ is not Martin-L\"of random, which indeed is a contradiction (see for example Downey and Hirschfeldt~\cite{DowneyHirschfeldtTA}). 

Since $\alpha$ is Martin-L\"of random and is left-c.e., we can apply the Ku$\check{\mathrm{c}}$era-Slaman theorem~\cite{KuceraS2001}. This theorem states that given a Martin-L\"of random left-c.e.\ real $\eta$, for any left-c.e.\ real $\xi$, there exists a constant $d$ and a partial recursive function $\varphi$ such that for every rational $q<\eta$, $\varphi(q)$ is defined and $\xi-\varphi(q) < 2^d(\eta -q)$. We will use this fact for $\eta=\alpha$ and $\xi=\Omega$ and also call~$d$ and~$\varphi$ the associated constant and partial recursive function. 

Now, let $c$ be a large integer (to be specified later). 
Suppose also that $\alpha \uh k$ is given for some~$k$. Since $\alpha - (\alpha \uh k) < 2^{-k}$, by the Ku$\check{\mathrm{c}}$era-Slaman theorem:
$$
\Omega - \varphi(\alpha \uh k) < 2^{-k+d}
$$
Thus, from $\alpha \uh k$, one can effectively compute some $s(k)$ such that
\begin{equation*}
\sum_{n>s(k)} 2^{-K(n)} \leq 2^{-k+d}
\end{equation*}
If $k$ is large enough, then $s(k)$ is large enough and hence $n>s(k) \Rightarrow K(n) \leq f(n)-c-d$ (this because~$f$ is not a Solovay function). Thus, for all~$k$ large enough:
$$
\sum_{n>s(k)} 2^{-f(n)} \leq 2^{-c-d} \sum_{n>s(k)} 2^{-K(n)} \leq 2^{-c-d} \cdot 2^{-k+d} \leq 2^{-k-c}
$$
This tells us that for~$k$ large enough, knowing $\alpha \uh k$ suffices to compute $s(k)$ and then (by the above inequality) effectively compute an approximation of $\alpha$ by at most $2^{-k-c}$. In other words, $\alpha \uh (k+c)$ can be computed from $\alpha \uh k$ and~$c$. Therefore, for all~$k$ large enough:
$$
K(\alpha \uh (k+c)) \leq K(\alpha \uh k, c) +O(1) \leq K(\alpha \uh k)+2\log c +O(1)
$$
The constant~$d$ is fixed, and $c$ can be taken arbitrarily large. Choose $c$ such that the expression $2\log c+O(1)$ in the above inequality is smaller than $c/2$. Then, for all~$k$ large enough,
$$
K(\alpha \uh (k+c)) \leq K(\alpha \uh k)+c/2
$$
An easy induction then shows that $K(\alpha \uh k) \leq O(k/2)$, contradicting the fact that $\alpha$ is random. 
\end{proof}

An interesting corollary of this theorem is that there are nondecreasing Solovay functions (which is not really obvious from the definition). To see that it is the case, it suffices to take a computable sequence $(r_n)_{n \in \N}$ of rational numbers such that every $r_n$ is a negative power of~$2$, the $r_n$ are nonincreasing and $\sum_n r_n$ is a Martin-L\"of random number (it is very easy to see that such sequences exist). Then, take $f(n)=- \log (r_n)$ for all~$n$. The function~$f$ is computable, nondecreasing and by Theorem~\ref{thm:mlr-sum} is a Solovay function.

\section{Solovay functions and Martin-L\"of randomness}

One of the most fundamental theorems of algorithmic randomness is the Levin-Schnorr theorem, proven independently by Levin and Schnorr in the 1970's. It characterizes Martin-L\"of random sequences by the prefix-free Kolmogorov complexity of their initial segments. More precisely, a sequence $\alpha \in \cs$ is Martin-L\"of random if and only if
$$
K(\alpha \uh n) \geq n-O(1)
$$
This theorem left open a fundamental question: is there a similar characterization of Martin-L\"of randomness in terms of plain Kolmogorov complexity?

\subsection{The Miller-Yu theorem}

This question remained open for almost three decades. It was finally answered positively in a recent paper of  Miller and Yu~\cite{MillerY2008}. 

\begin{thm}[Miller and Yu\footnote{G\'acs~\cite{Gacs1980} proved the equivalence $(i) \Leftrightarrow (ii)$ of Theorem~\ref{thm:milleryu1}}~\cite{MillerY2008}]\label{thm:milleryu1}
Let $\alpha \in \cs$. The following are equivalent:\\
(i) $\alpha$ is Martin-L\"of random.\\
(ii) $C(\alpha \uh n) \geq n-K(n)-O(1)$.\\
(iii) For all functions~$f \in \kub$, $C(\alpha \uh n) \geq n-f(n)-O(1)$. 
\end{thm}

Remarkably, Miller and Yu showed that in the item $(iii)$ above, the ``for all~$f$'' part can be replaced by a  \emph{single} function:

\begin{thm}[Miller and Yu~\cite{MillerY2008}]\label{thm:milleryu2}
There exists a function $g \in \kub$ such that for all $\alpha \in \cs$:
\begin{equation}\label{eq:milleryu}
\text{$\alpha$ is Martin-L\"of random} ~\Leftrightarrow ~C(\alpha \uh n) \geq n - g(n) - O(1)
\end{equation} 
\end{thm}

Informally, the function~$g \in \kub$ in this last proposition is a ``good'' upper bound of~$K$, in the sense that it is close enough to~$K$ to make possible the replacement of~$K$ by~$g$ in the equivalence $(i) \Leftrightarrow (ii)$ of Theorem~\ref{thm:milleryu1}. This reminds us of the Solovay functions which are also ``good'' upper bounds of~$K$ in their own way. And indeed, the function~$g$ constructed by Miller and Yu to make the equivalence~(\ref{eq:milleryu}) true is a Solovay function. We will show that this is not a coincidence, as \emph{all} functions~$g$ satisfying~(\ref{eq:milleryu}) are Solovay functions. But before that, we state a related theorem:

\begin{thm}[Bienvenu and Merkle~\cite{BienvenuM2007b}]\label{thm:dec-mach-ml}
A sequence $\alpha$ is Martin-L\"of random if and only if for all~$f \in \kub$, $f(\alpha \uh n) \geq n -O(1)$. Moreover, there exists a unique function $g \in \kub$ such that
\begin{equation}\label{eq:mlr-decidable}
\text{$\alpha$ is Martin-L\"of random} ~\Leftrightarrow ~ g(\alpha \uh n) \geq n  - O(1)
\end{equation} 
\end{thm}

We will prove:

\begin{thm}\label{thm:miller-yu-solf}
Any function~$g$ satisfying the equivalence~(\ref{eq:milleryu}) of Theorem~\ref{thm:milleryu2} is a Solovay function. The same is true for any function~$g$ satisfying the equivalence~(\ref{eq:mlr-decidable}) of Theorem~\ref{thm:dec-mach-ml}. 
\end{thm}

In order to prove this theorem, we show that in both characterizations of Martin-L\"of randomness ($K(\alpha \uh n) \geq n-O(1)$ and $C(\alpha \uh n) \geq n - K(n) - O(1)$) the lower bound on complexity is very sharp, that is there is no ``gap phenomenon''. 

\subsection{A ``no-gap'' theorem for randomness}

Chaitin~\cite{Chaitin1987} proved an alternative characterization of Martin-L\"of randomness: $\alpha \in \cs$ is Martin-L\"of random if and only if $K(\alpha \uh n)-n$ tends to infinity. Together with the Levin-Schnorr characterization, this shows a dichotomy: given a sequence $\alpha \in \cs$, either $\alpha$ is not Martin-L\"of random, in which case $K(\alpha \uh n)-n$ takes arbitrarily large negative values, or~$\alpha$ is Martin-L\"of random, in which case $K(\alpha \uh n)-n$ tends to $+\infty$. This means for example that there is no sequence $\alpha \in \cs$ such that $K(\alpha \uh n) = n +O(1)$. One may ask whether this dichotomy is due to a gap phenomenon, that is: is there a function~$h$ that tends to infinity, such that for every Martin-L\"of random sequence $\alpha$, $K(\alpha \uh n) \geq n+h(n)-O(1)$? Similarly, is there a function~$h'$ that tends to infinity such that for every sequence~$\alpha$, $K(\alpha \uh n) \geq n-h'(n)-O(1)$ implies that~$\alpha$ is Martin-L\"of random? We answer both these questions (and their plain complexity counterpart) negatively.

\begin{thm}\label{thm:no-gap}
There exists no function~$h:\N \rightarrow \N$ (computable or not) which tends to infinity and such that
$$
K(\alpha \uh n) \geq n -h(n) - O(1)
$$
is a sufficient condition for $\alpha$ to be Martin-L\"of random (in fact, not even for $\alpha$ to be Church stochastic). \\

\noindent Similarly, there is no function~$h:\N \rightarrow \N$  which tends to infinity and such that
$$
C(\alpha \uh n) \geq n - K(n)-h(n) - O(1)
$$
is a sufficient condition for $\alpha$ to be Martin-L\"of random (in fact, not even for~$\alpha$ to be Church stochastic).
\end{thm}

\begin{proof}

First, notice that  since we want to prove this for \emph{any} function that tends to infinity, we can restrict our attention to the nondecreasing ones. Indeed, if~$h$ is a function that tends to infinity, the function $$\tilde{h}(n)=\min\{f(k) \mid k \geq n\}$$ also tends to infinity and $\tilde{h} \leq h$.\\

Now, assume we are in the simple case where the function~$h$ is nondecreasing and computable. A standard technique to get a non-random binary sequence $\beta$ such that $K(\beta \uh n) \geq n-h(n)-O(1)$ is the following: take a Martin-L\"of random sequence $\alpha$, and insert zeroes into $\alpha$ in positions $h^{-1}(0),h^{-1}(1),h^{-1}(2),\ldots$. It is easy to see that the resulting sequence $\beta$ is not Martin-L\"of random (indeed, not even Church stochastic), and that the Kolmogorov complexity of its initial segments is as desired. This approach was refined by Merkle et al.~\cite{MerkleMNRS2006} where the authors used an insertion of zeroes on a co-c.e.\ set of positions in order to construct a left-c.e. sequence $\beta$ that is not Mises-Wald-Church stochastic, but has initial segments of very high complexity.\\

Of course, the problem here is that the function~$h$ in the hypothesis may be non-computable, and in particular may grow slower than any computable nondecreasing function. In that case, the direct construction we just described does not necessarily work: indeed, inserting zeroes at a noncomputable set of positions may not affect  the complexity of $\alpha$. To overcome this problem, we invoke the Ku$\check{\mathrm{c}}$era-G\'acs theorem (see  Ku$\check{\mathrm{c}}$era~\cite{Kucera1985}, G\'acs~\cite{Gacs1986}, or Merkle and Mihailovic~\cite{MerkleM2004}). This theorem states that any subset of $\N$ (or function from $\N$ to $\N$) is Turing-reducible to a Martin-L\"of random sequence.  Hence, instead of choosing \emph{any} Martin-L\"of sequence $\alpha$, we pick one that computes the function $h^{-1}$ and then insert zeroes into $\alpha$ at positions $h^{-1}(0),h^{-1}(1),\ldots$. Intuitively, the resulting sequence $\beta$ should not be random, as the bits of $\alpha$ can be used to compute the places where the zeroes have been inserted. This intuition however is not quite correct, as inserting the zeroes may destroy the Turing reduction $\Phi$ from $\alpha$ to $h^{-1}$. In other words, looking at $\beta$, we may not be able to distinguish the bits of $\alpha$ from the inserted zeroes.\\

The trick to solve this last problem is to delay the insertion of the zeroes to ``give enough time'' to the reduction $\Phi$ to compute the positions of the inserted zeroes. More precisely, we insert the $k$-th zero in position $n_k=h^{-1}(k)+t(k)$ where $t(k)$ is the time needed by $\Phi$ to compute $h^{-1}(k)$ from $\alpha$. This way, $n_k$ is computable from $\alpha \uh n_k$ in time at most $n_k$. From this, it is not too hard to construct a computable selection rule that selects precisely the inserted zeroes, witnessing that $\beta$ is not Church stochastic (hence not Martin-L\"of random). Moreover, since the ``insertion delay'' only makes the inserted zeroes more sparse, we have $K(\beta \uh n)  \geq n-h(n)-O(1)$. And similarly, since $\alpha$ is Martin-L\"of random, we have by the Miller-Yu theorem: $C(\alpha \uh{n}) \geq n -K(n)-h(n)-O(1)$.\\

The formal details are as follows. Let $h$ be a nondecreasing function. By the Ku$\check{\mathrm{c}}$era-G\'acs theorem, let $\alpha$ be a Martin-L\"of random sequence and $\Phi$ be a Turing functional such that $\Phi^\alpha(n)=h^{-1}(n)$ for all~$n$. Let~$t(n)$ be the computation time of $\Phi^\alpha(n)$ (we can assume that~$t$ is a nondecreasing function). Let $\beta \in \cs$ be the sequence obtained by inserting zeroes into~$\alpha$ in positions $h^{-1}(n)+t(n)$. To show that $\beta$ is not Church stochastic, we construct a (total) computable selection rule that filters the inserted zeroes from~$\beta$. Let $S$ be the selection rule that works as follows on a given sequence $\xi \in \cs$. We proceed by induction; we call $k_n$ the number of bits selected by $S$ from $\xi \uh n$ and $x_n$ the prefix $\xi \uh n$ of $\xi$ from which these $k_n$ bits are deleted ($x_0$ is thus the empty string, and $k_0=0$). 

At stage~$n+1$, having already read $\xi \uh n$, $S$ computes $\Phi^{x_n}_n(k_n)$. If the computation halts after $s$ steps, $S$ checks whether $\Phi^{x_n}_n(k_n)+s$ returns~$n$. If so, $S$ selects the $n$-th bit of $\xi(n)$ of $\xi$ and then sets $x_{n+1}=x_n$ and $k_{n+1}=k_n+1$. Otherwise, $S$ just reads the bit $\xi(n)$, and sets $x_{n+1}=x_n\xi(n)$ and $k_{n+1}=k_n$.\\

It is clear that $S$ is a total computable selection rule. Now suppose that we run it on~$\beta$. We argue that $S$ selects exactly the zeroes that have been inserted into~$\alpha$ to get~$\beta$.  We prove this by induction. If $S$ has already selected from $\beta$ the first~$i$ inserted zeroes, then the next selected bit is the bit in position $n=\Phi^{x_n}(k_n)+s$ where $\Phi^{x_n}(k_n)$ is computed in $s$ steps. But since the selected bits are exactly the zeroes that were inserted in $\alpha$, we have $k_n=i$ and $x_n=\alpha \uh n-i$, and thus $s$ is the computation time of $\Phi^{x_n}(k_n)=\Phi^{\alpha \uh n-i}(i)$, which we called~$t(i)$. And by definition of $\Phi$, $\Phi^{\alpha \uh n-i}(i)=h^{-1}(i)$. Therefore, $n=h^{-1}(i)+t(i)$, i.e.\ the selected bit was an inserted zero. This proves that $S$ only selects bits that belong to the zeroes that were inserted into $\alpha$. Conversely, we need to prove that all such bits are indeed selected by~$S$. Let $i \in \N$. The $i+1$-th inserted zero is in position $n=h^{-1}(i)+t(i)$. At stage~$n$, we have by the induction hypothesis $x_n=\alpha \uh n-i$ and $k_n=i$. Thus, $\Phi^{x_n}_n(k_n)=\Phi^{\alpha \uh t(i)+h^{-1}(i)-i}_{h^{-1}(i)+t(i)}(i)$, which has to halt because both quantities $t(i)+h^{-1}(i)-i$ and $h^{-1}(i)+t(i)$ are greater than $t(i)$, which is the computation time of $\Phi^\alpha(i)$. Thus the bit in position~$n$ is indeed selected. Therefore, $S$ satisfies the desired properties, and witnesses the fact that $\beta$ is not Church stochastic.

Finally, for all~$n$, calling $i$ the number of inserted zeroes in $\beta \uh n$, we easily see that $\beta \uh n$ and $\alpha \uh n-i$ can each be computed from the other one (by insertion or deletion of zeroes). Thus:
$K(\beta \uh n)=K(\alpha \uh n-i) \geq n-i$ (since $\alpha$ is Martin-L\"of random). And by definition of the positions where the zeroes are inserted, we have $n \geq h^{-1}(i-1)+t(i-1)$, hence $i \leq h(n)+O(1)$. Therefore:
$$
K(\beta \uh n) \geq n-i \geq n-h(n)+O(1)
$$
for all~$n$. This completes the proof.
\end{proof}

As a consequence of the construction performed in this proof, we get the dual version of Theorem~\ref{thm:no-gap}:

\begin{prop}
There exists no function~$h:\N \rightarrow \N$ (computable or not) which tends to infinity and such that
$$
K(\alpha \uh n) \geq n + h(n) - O(1)
$$
is a necessary condition for $\alpha$ to be Martin-L\"of random. \\

\noindent Similarly, there is no function~$h:\N \rightarrow \N$  which tends to infinity and such that
$$
C(\alpha \uh n) \geq n - K(h) + h(n) - O(1)
$$
is a necessary condition for $\alpha$ to be Martin-L\"of random.
\end{prop} 

\begin{proof}
Suppose for the sake of contradiction that there exists a function~$h'$ which tends to infinity and such that
$K(\alpha \uh n) \geq n + h'(n) - O(1)$ is a necessary condition for $\alpha$ to be Martin-L\"of random. Once again, we can assume that~$h'$ is non-decreasing. Then, we perform the exact same construction as in the proof of Theorem~\ref{thm:no-gap} for a given function~$h$. Then, at the end of proof, when evaluating the complexity of $\beta$, we have $K(\beta \uh n)=K(\alpha \uh n-i)+O(1)$, with $i \leq h(n)+O(1)$, and since $\alpha$ is Martin-L\"of random, $K(\alpha \uh n-i) \geq (n-i)+h'(n-i)-O(1)$. It follows that
$$
K(\beta \uh n) \geq n-h(n)+h'(n-h(n))-O(1)
$$
Thus, if we take $h$ to be sufficiently slow growing (for example $h(n)=\log(h'(n))$), we have $K(\beta \uh n) \geq n-O(1)$ for all~$n$. This is a contradiction since by the Levin-Schnorr theorem, this would imply that $\beta$ is Martin-L\"of random, which it is not by construction. The proof of the second part of the proposition is almost identical. 
\end{proof}

Theorem~\ref{thm:miller-yu-solf} now easily follows:

\begin{proof}[Proof (of Theorem~\ref{thm:miller-yu-solf})]
Let $g$ be a function satisfying the equivalence~(\ref{eq:milleryu}) of Theorem~\ref{thm:milleryu2}. Suppose that~$g$ is not a Solovay function. This means, by definition, that $h(n)=g(n)-K(n)$ tends to infinity. Then, we can rewrite the equivalence~(\ref{eq:milleryu}) as:
$$
\text{$\alpha$ is Martin-L\"of random} ~\Leftrightarrow ~C(\alpha \uh n) \geq n - K(n) - h(n) - O(1)
$$
which contradicts Theorem~\ref{thm:no-gap}. Similarly, if a function $g$ satisfies the condition~(\ref{eq:mlr-decidable}) of Theorem~\ref{thm:dec-mach-ml}, and is such that $h(n)=g(n)-K(n)$ tends to infinity, then for all $\alpha \in \cs$, $\alpha$ is Martin-L\"of random if and only if $K(\alpha \uh n) \geq n - h(n)$, contradicting Theorem~\ref{thm:no-gap}.
\end{proof}

The consequences of Theorem~\ref{thm:no-gap} go beyond its applications to Solovay functions. For example, it gives an alternative proof of the fact that Schnorr randomness does not imply Church stochasticity (a result originaly proven by Wang~\cite{Wang1999}). Indeed, it is rather well-known that if $h$ tends to infinity slower than any computable nondecreasing function, then the condition $K(\alpha \uh n) \geq n - h(n) -O(1)$ is sufficient for $\alpha$ to be Schnorr random (see for example Bienvenu and Merkle~\cite{BienvenuM2007b}), whereas we just saw that it was not sufficient for~$\alpha$ to be Church stochastic.\\

One can also adapt the proof of Theorem~\ref{thm:no-gap} to separate Church stochasticity from Schnorr randomness within the left-c.e.\ reals. Informally, this is done as follows. Take a left-c.e.\  Martin-L\"of random sequence $\alpha \in \cs$. Call~$t(n)$ the settling time of~$\alpha \uh n$, i.e.\ given a computable nondecreasing sequence $(q_s)_{s \in \N}$ that converges to~$\alpha$, $t(n)$ is the smallest~$s$ such that $|\alpha-q_s|<2^{-n}$. It is easy to see that~$t$ is enumerable from below. Thus, the sequence~$\beta \in \cs$ which we obtain from~$\alpha$ by inserting zeroes in positions $t(0)<t(1)<\ldots$ is left-c.e.\, and for the same reason as above, is not Church stochastic. And the same kind of computation as above shows that $K(\beta \uh n) \geq n - t^{-1}(n)-O(1)$. Since it can easily be shown that~$t$ grows faster than any computable function, it follows that $t^{-1}$ tends to infinity more slower than any nondecreasing unbounded computable function. Thus, $\beta$ is not Church stochastic. This improves a result of Merkle et al.~\cite{MerkleMNRS2006} (Theorem 26), who proved an equivalent fact for a weaker notion of stochasticity. For details on that result, see Bienvenu~\cite{BienvenuPhD-eng}. 

\section{Solovay functions and triviality notions}

A very successful line of research in algorithmic randomness over the last years concerns triviality and lowness notions. Informally, a sequence $\alpha \in \cs$ is trivial if its Kolmogorov complexity is minimal or quasi-minimal, while a sequence $\alpha$ is low for randomness if it has little computation power, i.e.\ if relativizing the definition of random sequences to the oracle~$\alpha$ does not change the class of random sequences. Perhaps the most important result in this direction was given by Nies~\cite{Nies2005}: a sequence $\alpha \in \cs$ is low for Martin-L\"of randomness (i.e.\ Martin-L\"of randomness relativized to~$\alpha$ coincides with standard Martin-L\"of randomness) if and only if $\alpha$ is K-trivial (i.e.\ $K(\alpha \uh n) \leq K(n)+O(1)$). Other interesting notions of triviality have been studied, like Schnorr triviality, introduced by Downey et al.~\cite{DowneyGL2004}: a sequence~$\alpha$ is Schnorr trivial if for every prefix-free machine~$M$ whose domain has measure~$1$, there exists a machine~$M'$ whose domain also has measure~$1$, and such that $K_{M'}(\alpha \uh n) \leq K_M(n)+O(1)$. This notion was extensively studied by Franklin~\cite{Franklin2008, Franklin2008b}.\\

In the same spirit, we can consider the class of sequences~$\alpha$ such that for all computable upper bounds~$f$ of~$K$, there exists a computable upper bound~$f'$ of~$K$ such that $f'(\alpha \uh n) \leq f(n)+O(1)$. However, because of the existence of Solovay functions, only computable sequences have this property.

\begin{prop}\label{prop:solovay-triv}
Let $\alpha \in \cs$. Suppose that for all $f \in \kub$, there exists $f' \in \kub$ such that $$f'(\alpha \uh n) \leq f(n)+O(1)$$
Then $\alpha$ is computable. 
\end{prop}
  
 To prove this proposition, we need the following lemma: 
 
 \begin{lemma}[Chaitin~\cite{Chaitin1976}]\label{lem:k-counting}
 For every $n,c \in \N$:
 $$
 \#\left\{w \in \fs \mid |w|=n~\wedge~K(w) \leq K(n)+c\right\} \leq 2^{c+O(1)}
 $$
 where the $O(1)$ term does not depend on $n$ or $c$. 
 \end{lemma}
 
 \begin{proof}[Proof (of Proposition~\ref{prop:solovay-triv})]
 Let $\alpha \in \cs$ satisfy the hypothesis of the proposition. Let $f$ be a Solovay function. By the assumption on~$\alpha$, there is a function $f' \in \kub$ and a constant $c$ such that $f'(\alpha \uh n) \leq f(n)+c$ for all~$n$. Let $d$ be a constant such that $K \leq f'+d$. Since~$f$ is a Solovay function, there exists a constant $e$ such that $f(n) \leq K(n)+e$ for infinitely many~$n$. For any such~$n$, we have:
$$
  \#\left\{w \in \fs \mid |w|=n~\wedge~f'(w) \leq f(n)+c\right\}
$$
$$  
\leq \#\left\{w \in \fs \mid |w|=n~\wedge~K(w) \leq K(n)+c+d+e\right\}
$$
$$
\leq 2^{c+d+e+O(1)}
$$
(the last inequality comes from Lemma~\ref{lem:k-counting}). From this, we see that the $\Pi^0_1$ class
$$
\{\xi \in \cs \mid \forall n ~ f'(\xi \uh n) \leq f(\xi \uh n)+c\}
$$
to which $\alpha$ belongs, has only finitely many elements (at most $2^{c+d+e+O(1)})$, hence all these elements are computable.
 \end{proof} 
  
 Another thing we can do is to study a weakened version of K-triviality: we consider the class of sequences~$\alpha$ such that for any $f \in \kub$, $K(\alpha \uh  n) \leq f(n)+O(1)$. As we shall now see, this is equivalent to K-triviality, hence we obtain an analogue of the Miller-Yu theorem for K-triviality. 
 
\begin{thm}\label{thm:k-triv-kub}
Let $\alpha \in \cs$. Then, $\alpha$ is K-trivial if and only if for all functions $f \in \kub$, $K(\alpha \uh n) \leq f(n)+O(1)$. Moreover, there exists a unique function $g \in \kub$ such that for all $\alpha \in \cs$:
\begin{equation}\label{eq:k-triv-solf}
 \text{$\alpha$ is K-trivial} \Leftrightarrow  K(\alpha \uh n) \leq g(n)+O(1)
\end{equation}
\end{thm}

\begin{proof}
By Lemma~\ref{lem:kub-sum}, it is obvious that any K-trivial $\alpha$ satisfies $K(\alpha \uh n) \leq f(n)+O(1)$ for all $f \in \kub$. Thus, all we have to do to prove this theorem is to construct a function $g$ such that the implication ``$\Leftarrow$'' of equation (\ref{eq:k-triv-solf}) holds. In fact, we just take for $g$ the function~$f$ constructed in the proof of Theorem~\ref{thm:existence-solf}.\\ 

Let then $\alpha$ be a sequence such that $K(\alpha \uh n) \leq g(n)+c$ for some constant~$c$ and all~$n$. As usual, we prove that $\alpha$ is K-trivial by building a c.e.\ set $L$ of pairs $(w_i,k_i)_{i \in \N}$ (with $w_i \in \fs$ and $k_i \in \N$) such that $\sum_i 2^{-k_i} < +\infty$ and for all~$n$, some pair $(\alpha \uh n, K(n)+O(1))$ belongs to $L$.\\

Let $n$ be a fixed integer. We describe the strategy to enumerate strings of length~$n$ into~$L$. We proceed by stages. At stage~$s$, we look at the value of $K_s(n)$ , and work under the assumption that $K_s(n)=K(n)$ (this assumption might turn out to be incorrect, we shall see below what to do when this happens). We then effectively find a $\U$-program $p$ of length at most $K_s(n)$ such that $\U(p)=n$. By definition of $g$, if we call $t$ the computation time of $\U(p)$, we have $g(\langle n,p,t \rangle)=|p| \leq K_s(n)$ (by definition of the function~$g$), which, under the assumption $K_s(n)=K(n)$ implies $K(\langle n,p,t \rangle) = K(n)+O(1) = g(\langle n,p,t \rangle)+O(1)$. In other words, at every stage $s$, we can find a witness $m_s=\langle n,p,t \rangle$ such that $K(m_s) = g(m_s)+O(1)$, provided $K_s(n)=K(n)$.\\

Then, we enumerate all strings~$w$ of length $m_s$ such that $K(w) \leq g(m_s)+c$, and for each such string we find, we put $(w \uh n, K_s(n))$ into~$L$ (without repetitions). Under the assumption $K_s(n)=K(n)$, we have $g(m_s)=K(m_s)+O(1)$ hence by Lemma~\ref{lem:k-counting}, there are at most $d=2^{c+O(1)}$ different strings $w$ of length $m_s$ such that $K(w) \leq g(m_s)+c$, hence at most $d$ pairs of type $(w \uh n,K_s(n))$ enter~$L$.\\

As we noted above, we might realize at some point that the assumption $K_s(n)=K(n)$  is incorrect, i.e.\ there might exist a stage $s'>s$ such that $K_{s'}(n)<K_s(n)$. In this case, we simply compute a new witness $m_{s'}$ and restart the strategy. However, the false assumption $g(m_s)=K(m_s)+O(1)$ may have caused us to enumerate many strings $w$ of length $m_s$ such that $K(w) \leq g(m_s)+c$ hence many pairs $(w \uh n,K_s(n))$ may enter~$L$. We avoid this situation by only allowing $d$ such pairs to enter $L$. Indeed, if more than~$d$ such pairs ask to enter~$L$, we immediately know that the assumption~$K_s(n)=K(n)$ is incorrect, hence we can stop acting and simply wait for a stage $s'$ such that $K_{s'}(n)<K_{s}(n)$ and only then restart the strategy.\\

It remains to be verified that this strategy works, i.e.\ that the set~$L$ has the desired properties. For a fixed~$n$, and any $k \geq K(n)$, by construction of~$L$, there are at most~$d$ pairs of type $(w \uh n,k)$ in~$L$. Thus, the total measure of the domain of~$L$ is at most
$$
\sum_n \sum_{k \geq K(n)} d \cdot 2^{-k} = \sum_n d \cdot 2^{-K(n)+1} \leq 2d
$$
hence is finite. Finally, for a given~$n$, at some stage~$t$ we do have $K_t(n)=K(n)$. We then have $g(m_t)=K(m_t)+O(1)$ hence for \emph{all} strings $w$ of length $m_t$ satisfying $K(w \uh m_t) \leq g(m_t)+c$, the pair $(w�uh n, K_t(n))$ is enumerated into $L$ (the restriction that at most $d$ such pairs can enter~$L$ is not an actual restriction when $g(m_t)=K(m_t)+O(1)$). By definition of $\alpha$, $K(\alpha \uh m_t) \leq g(m_t)+c$, hence by assumption $(\alpha \uh n, K_t(n))=(\alpha \uh n,K(n))$ is enumerated into~$L$. This completes the proof. 
\end{proof}

We would like to end this paper with two questions. 

\begin{question}
Does \emph{any} Solovay function~$g$ make the equivalence~(\ref{eq:milleryu}) of Miller-Yu's theorem true?
\end{question}

\begin{question}
Is any computable function~$g$ satisfying the equivalence~(\ref{eq:k-triv-solf}) of Theorem~\ref{thm:k-triv-kub} necessarily a Solovay function?
\end{question}

Note that one cannot invoke a ``no-gap'' theorem to answer the second question, as it was noted by Csima and Montalb\'an~\cite{CsimaM2005} that there \emph{is} a nondecreasing unbounded function~$h$ such that $K(\alpha \uh n) \leq K(n)+h(n)+O(1)$ implies that~$\alpha$ is K-trivial. 

\section*{Acknowledgement}
\noindent The first author is grateful to Serge Grigorieff for very useful discussions on this work. The authors also thank the anonymous referees of this paper for helpful suggestions. 


\bibliographystyle{plain}
\bibliography{biblio}

\end{document}